\documentclass[conference,9pt,english,twocolumn]{IEEEtran}
\usepackage{lmodern}
\usepackage{amssymb,amsmath}
\usepackage{ifxetex,ifluatex}
\usepackage{fixltx2e} 
\ifnum 0\ifxetex 1\fi\ifluatex 1\fi=0 
  \usepackage[T1]{fontenc}
  \usepackage[utf8]{inputenc}
\else 
  \ifxetex
    \usepackage{mathspec}
  \else
    \usepackage{fontspec}
  \fi
  \defaultfontfeatures{Ligatures=TeX,Scale=MatchLowercase}
\fi
\IfFileExists{upquote.sty}{\usepackage{upquote}}{}
\IfFileExists{microtype.sty}{%
\usepackage{microtype}
\UseMicrotypeSet[protrusion]{basicmath} 
}{}
\usepackage[unicode=true]{hyperref}
\usepackage[2023, cc-by, pages={5}{6}, url=https://osda.ws/r/ifIbv]{osda}
\hypersetup{
            pdftitle={Newad: A register map automation tool for Verilog},
            pdfkeywords={Verilog, VerilogHDL, VHDL, Regsiter
map, Automation, Address map, Code generation},
            pdfborder={0 0 0},
            breaklinks=true}
\ifnum 0\ifxetex 1\fi\ifluatex 1\fi=0 
  \usepackage[shorthands=off,main=english]{babel}
\else
  \usepackage{polyglossia}
  \setmainlanguage[]{}
\fi
\usepackage[]{biblatex}
\IfFileExists{parskip.sty}{%
\usepackage{parskip}
}{
\setlength{\parindent}{0pt}
\setlength{\parskip}{6pt plus 2pt minus 1pt}
}
\ifx\paragraph\undefined\else
\let\oldparagraph\paragraph
\renewcommand{\paragraph}[1]{\oldparagraph{#1}\mbox{}}
\fi
\ifx\subparagraph\undefined\else
\let\oldsubparagraph\subparagraph
\renewcommand{\subparagraph}[1]{\oldsubparagraph{#1}\mbox{}}
\fi

\makeatletter
\def\fps@figure{htbp}
\makeatother

\title{Newad: A register map automation tool for Verilog}

\author{
            \IEEEauthorblockN{Vamsi K Vytla}
        \IEEEauthorblockA{%
            Lawrence Berkeley National Lab \\
            Berkeley, California, USA \\
            vkvytla@lbl.gov \\}
         \and
            \IEEEauthorblockN{Larry Doolittle}
        \IEEEauthorblockA{%
            Lawrence Berkeley National Lab \\
            Berkeley, California, USA \\
            lrdoolittle@lbl.gov \\}
        }

\date{}

\begin{document}
\maketitle
\begin{abstract}
Large scale scientific instrumentation-and-control FPGA gateware designs
have numerous run-time settable parameters. These can be used either for
user-level control or by automated processes (e.g., calibration). The
number of such parameters in a single design can reach on the order of
1000, and keeps evolving as the gateware and its functionality evolves.
One must keep track of which module the registers belong to, where the
registers need to be decoded, and how to express the properties (or even
semantics) of the register to the next level of user or software. Note,
the registers maybe embedded anywhere throughout the module hierarchy.
Purely manual handling of these tasks by HDL developers is considered
burdensome and error-prone at this scale. Typically these registers are
writable via an on-chip bus, vaguely VME-like, that is controlled by an
on-chip or off-chip CPU. There have been several attempts in the
community to address this task at different levels. However, we have
found no tool that is able to generate a register map, generate decoders
and encoders with minimal overhead to the developer. So, here we present
a tool that scours native HDL source files and looks for specific
language-supported attributes and automatically generates a register map
and bus decoders, respecting multiple clock domains, and presents a JSON
file to the network that maps register names to addresses.
\end{abstract}


\hypertarget{overview}{%
\section{Overview}\label{overview}}

The document describes \texttt{newad} \autocite{newad}, our address map
code generation tool for VerilogHDL designs. Initially, we describe the
motivation for developing such a tool, followed by commenting on some
design considerations and the design itself. Finally, we describe the
tool in some detail, ending with a conclusion of what we think we
achieved and what can be done in the future.

\hypertarget{introduction}{%
\section{Introduction}\label{introduction}}

\hypertarget{motivation}{%
\subsection{Motivation}\label{motivation}}

Larger experiments demand larger register map sets. This section
elaborates a few reasons of why develop a register map code generation
tool.

\begin{enumerate}
\def\labelenumi{\arabic{enumi}.}
\item
  Repetitive and error prone: With the FPGA LUT count ever increasing,
  gateware designs today are larger than ever. This invariably leads to
  a large set of registers/register maps that are managed from software.
  When lacking automation, the developer manually tracks registers
  during development or maintenance cycles where new registers are
  created, removed, and various attributes about them are constantly
  changing. This process let alone repetitive, can also be error prone
  demanding a register map code generation engine.
\item
  Bus decoder: Manually decoding several register values from a system
  bus (address and data lines) is a tedious task. In addition to the
  decoding, developer needs to decide in which module should she
  instantiate the decoder. This is especially tedious in an evolving
  design. We propose an engine that instantiates the bus decoder at a
  requested level at pre-compile time. The developer may optionally
  choose, to provide some register values as readback to the user of the
  design, depending on the resources available. It is preferable to
  manage such register decoding/encoding in an automated way.
\item
  Register attributes: IEEE register attributes, defined
  \autocite{attributes}. Registers have several attributes that a user
  is expected to be aware of such as it's size, read-only nature, bit
  descriptions, clock domain, a physical formula it relates to, and a
  human readable description (documentation). The goal is to
  automatically gather such attributes and present it to both the end
  user (in a readable format like PDF) and to other software (in a
  machine comprehensible format like a c-Header/JSON).
\item
  Documentation generation: Developers don't need to write documentation
  outside the source code. All modern languages and tools allow for
  this. Given code for registers and memory maps can be generated from
  source code, the documentation about them should be generated as well.
\end{enumerate}

With the above motivation in mind, and working with VerilogHDL over the
years, we have written a developer tool called \texttt{newad} that
parses Verilog code pre-compile time, and accomplishes many of the above
tasks.

\hypertarget{design}{%
\section{Design}\label{design}}

The register map automation started out of necessity as the designs were
steadily increasing in size. Our initial attempt at code generation, was
achieved by the developer placing ``magic''-comments throughout a
Verilog project. \texttt{newad} looked for these comments and generated
the necessary register map, and bus decoders. Relying on comments from
source files isn't a great strategy. Knowing this we looked around for
other tools that have solved the problem

We came across \texttt{Cheby} \autocite{cheby} developed by engineers at
CERN (European Organization for Nuclear Research). Cheby works by
requiring the developer to maintain a parallel file to each source file,
where the registers being used in the source file are described in YAML.
Cheby then source agnostically parses the register (YAML) file and is
capable of generating a C header/Documentation/HDL instantiation of the
YAML representation. This more or less solves the problems we have
discussed above, all the while introducing new ones.

A serious limitation to Cheby is the requirement that the user needs to
maintain a YAML file describing registers. This adds overhead to the
developer where now she needs to look into more than one file for the
truth. Readability is affected.

After continuing to look for other options the authors decided to
leverage using language attributes to encode information that was being
put inside magic-comments. Attributes make this a lot cleaner. Verilog
attributes are defined to be rather simple but clean and allow for
cleanly encoding lots of register related information.

To leverage the usage of attributes, and successfully parse them, it
made sense to use an existing Verilog parser. In order to accomplish
this, we looked at iverilog and Yosys parsers. Attribute parsing wasn't
sufficiently implemented in either of those open source parsers, and it
took a few contributions from the authors to add the necessary features.
After using both, we stuck with Yosys, for both technical and licensing
reasons.

\hypertarget{project-details}{%
\section{Project Details}\label{project-details}}

\hypertarget{registers}{%
\subsection{Registers}\label{registers}}

Registers maybe marked as ``automatic'' using an attribute in any module
of the design hierarchy. Registers are marked automatic at module
definition site, as shown in the code sample below.

A developer may also notify \texttt{newad} to generate additional
register logic, such as a write strobe for the register, a read strobe,
a signal that holds the write value only for single cycle, etc.

\begin{verbatim}
module prng(
        input clk,
        output [31:0] rnda,
        output [31:0] rndb,
        (* external *)
        input [0:0] run,
        (* external, signal_type="plus-we" *)
        input [31:0] iva,
        input iva_we,  // special trailing _we

);
\end{verbatim}

\hypertarget{module-instances}{%
\subsection{Module instances}\label{module-instances}}

If a module has any registers described as ``automatic'', it is expected
that signals/wires for the register need to be routed through its module
instantiation site. When the developer marks the instantiation site with
``automatic'', \texttt{newad} then generates a macro for that
instantiation site. The macro contains wires for the automatic registers
of to the module being instantiated.

\begin{verbatim}
(* lb_automatic *)
prng prng (
    .clk(clk),
    .rnda(rnda),
    .rndb(rndb)
    `AUTOMATIC_prng
);
\end{verbatim}

\hypertarget{verilog-header-files}{%
\subsection{Verilog header files}\label{verilog-header-files}}

\texttt{newad} is run as a pre-compile step. Once \texttt{newad} is run
the macros are populated inside generated header files. The developer is
expected to include these header files. This strictly keeps all
generated code away from source. We are still considering a scheme where
\texttt{newad} generates a single Verilog file with all modules it has
parsed and the generated code.

Continuing with the example above, for the sake of simplicity let's say
the prng module was included in a top-level file, known as station. The
following automatically generated header files can be included inside
station.v. First we will look at \texttt{station\_auto.vh} which is a
headerfile expected to be included with station.v.

\begin{verbatim}
// station_auto.vh
// parse_vfile_yosys  station.v
..
// module=prng instance=prng gvar=None gcnt=None
// parse_vfile_yosys :station.v ./prng.v
`define AUTOMATIC_prng .run(prng_run),\
    .iva(prng_iva),\
    .iva_we(prng_iva_we),\
..
\end{verbatim}

Second, \texttt{addr\_map\_station.vh} is a generated file that strictly
includes the address map that was generated by newad.

\begin{verbatim}
// addr_map_station.vh
`define LB_HI 14
..
 // prng_iva bw: 0, base_addr: 7203
`define HIT_prng_iva (lb_addr[`LB_HI:0]==7203)
 // prng_run bw: 0, base_addr: 7205
`define HIT_prng_run (lb_addr[`LB_HI:0]==7205)
\end{verbatim}

A JSON file is generated as an API for a top-level software to be able
to access the register info.

\begin{verbatim}
..
    "prng_iva": {
        "access": "rw",
        "addr_width": 0,
        "base_addr": 7203,
        "data_width": 32,
        "description": "",
        "sign": "unsigned"
    },
    "prng_run": {
        "access": "rw",
        "addr_width": 0,
        "base_addr": 7205,
        "data_width": 1,
        "description": "",
        "sign": "unsigned"
    }
..
\end{verbatim}

\hypertarget{decoder-generation}{%
\subsection{Decoder generation}\label{decoder-generation}}

Once newad is notified of a top-level Verilog file, and recursively
searches paths for the modules utilized in the design, \texttt{newad}
builds a tree with per-module register information. It can then generate
a bus decoder upon programmer's command. Generated code for the decoder
is embedded in a macro which is utilized at the intended module.

\hypertarget{conclusion}{%
\section{Conclusion}\label{conclusion}}

We believe large, complex projects need a robust scheme to manage their
register space. Languages like Verilog, VHDL, and bSystem Verilog that
can't cleanly express such semantics natively demand a tool like
\texttt{Cheby} or \texttt{newad}. \texttt{newad} initially emerged out
of necessity and evolved into something that is actively supporting
several gateware projects. We have seen at least one modern HDL language
such as (n)-migen (a Python based DSL for describing gateware), where
such register map generation is fully embedded into the language inside
a library.

We intend to release \texttt{newad} as a stand-alone package for
register map automation. Currently, it is embedded inside our framework
as a build tool
\href{https://github.com/BerkeleyLab/Bedrock/blob/newad_yosys_take1/build-tools/newad.py}{here}.
We believe such register map automation could be done for VHDL as well.
However, VHDL doesn't play very well with header files.

\texttt{newad} is an HDL-developer-centric approach to register space
management. It focuses on HDL readability and maintainability, and a
single-source-of-truth for generating register maps and their
documentation.

\printbibliography[title=References]

\end{document}